\documentclass[a4paper,12pt]{article}
\usepackage[centertags]{amsmath}
\usepackage{xcolor}
\usepackage{amsfonts}
\usepackage{pst-node}
\usepackage{tikz-cd} 
\usepackage{amssymb}
\usepackage{float}
\usepackage[left=1in,right=1in,top=1.2in, bottom=1in]{geometry}
\usepackage{amsthm}
\usepackage{multirow}

\newtheorem{theorem}{Theorem}[section]
\newtheorem{corollary}{Corollary}[section]
\newtheorem{proposition}{Proposition}[section]
\newtheorem{lemma}{Lemma}[section]
\newtheorem{example}{Example}[section]
\newtheorem{remark}{Remark}[section]
\newtheorem{definition}{Definition}[section]

\newcommand{\ZZ}{{\mathbb{Z}}}
\newcommand{\CC}{{\mathbb{C}}}
\newcommand{\F}{{\mathbb{F}}}
\newcommand{\Z}{{\mathbb{Z}}}

\newcommand{\vc}{{\mathbf{c}}}
\newcommand{\vw}{{\mathbf{w}}}
\newcommand{\vv}{{\mathbf{v}}}
\newcommand{\vu}{{\mathbf{u}}}

\newcommand{\FF}{{\mathbb{F}}}
\newcommand{\QQ}{{\mathbb{Q}}}

\newcommand{\fa}{{\mathfrak{a}}}

\newcommand{\fb}{{\mathfrak{b}}}
\newcommand{\fB}{{\mathfrak{B}}}
\newcommand{\fA}{{\mathfrak{A}}}
\newcommand{\fm}{{\mathfrak{m}}}
\newcommand{\cP}{{\cal{P}}}
\newcommand{\wcP}{{\widehat{\cP}}}

\newcommand{\zero}{{\mathbf{0}}}
\newcommand{\one}{{\mathbf{1}}}

\title{ Weight Enumerators From Equivalence Relations and MacWilliams Identities}
\author{
Steven Dougherty \\
Department of Mathematics \\
University of Scranton\\
USA\\
Cristina Fern\'{a}ndez-C\'{o}rdoba \\
Department of Information and Communication Engineering\\
Universitat Aut{\`o}noma de Barcelona \\
Spain\\
}

\begin{document}	
	\maketitle	
\footnotetext[1]{ This work has been partially supported by the Spanish MCIN under Grant PID2022-137924NB-I00 (AEI / 10.13039/501100011033 / FEDER, UE), and by the Catalan AGAUR under grant 2021 SGR 00643.}

	\begin{abstract}
 In this paper, we consider codes over finite fields, finite abelian groups, and finite Frobenius rings. For such codes, the complete weight enumerator and the Hamming weight enumerator serve as powerful tools.  These two types of weight enumerators satisfy the MacWilliams relations. We define the weight enumerator of a code with respect to an equivalence relation and determine in which cases the MacWilliams relations hold for this weight enumerator. We also study some weight enumerators for specific equivalence relations.
	\end{abstract}

{\bf Keywords:}
MacWilliams relations;\;  codes over chain rings; \;equivalence relations;\\ \;weight enumerators

\section{Introduction} 

One of the foundational results of algebraic coding theory is the MacWilliams relations, first appearing in Jessie MacWilliams' 1961 thesis in  \cite{thesis} and  later published in \cite{Mac2}.
Initially, the MacWilliams relations, which relate the weight enumerator of a code with the weight enumerator of its orthogonal,  were proven for linear codes over the binary field, but were later expanded to the complete weight enumerator and the Hamming weight enumerator for codes over arbitrary finite fields.  These theorems have been one of the most powerful tools in coding theory and help in the study of nearly every part of coding theory.   

Following a surge in the study of codes over finite rings, Jay Wood, studied the question: ``What is the largest class of rings for which the MacWilliams relations hold?"  The answer was given by Jay Wood in papers  \cite{Wood} and \cite{Wood2}.  Namely, he showed that Frobenius rings were the largest class of rings for which these relations hold and showed explicitly what these relations were in this case.

In this paper, we give the definition of new weight enumerators, and we determine the MacWilliams relations for these weight enumerators. 
In Section~\ref{due}, we give the necessary notation, definitions, and foundational results to prove our results. We also give basic definitions for chain rings, and some results related to codes over chain rings. In Section~\ref{tre}, we define the weight enumerator of a code with respect to a given equivalence relation. Then, we study some weight enumerators for specific equivalence relations.

% {\color{red} In Section~\ref{tre}, we define a weight enumerator for additive codes based on the order of the elements in the group and give the MacWilliams relations for them. In Section~\ref{subsection:CRW}, we define a weight enumerator for linear  codes  over a finite Frobenius ring based on ideals and give the MacWilliams relations for them. }

\section{Additive and Linear Codes} \label{due} 

\subsection{Additive codes over groups and rings and linear codes over  rings}

We begin by giving the standard definitions needed to describe additive and linear codes over a general alphabet. In the next definition, we define both additive and linear codes.

\begin{definition}
A code over any alphabet $A$ of length $n$ is a subset of $A^n$. 
\begin{itemize}
\item  If  $A=G$ is a finite abelian group, then an additive code over $G$ of length $n$ is an additive subgroup of $G^n.$ 
\item   If $A=R$ is a finite Frobenius ring, a linear code  over $R$ of length $n$ is a $R$-submodule of $R^n$. 
\end{itemize} 
\end{definition}  

We note that for additive codes, we are only interested in one operation, namely the operation of the group.  Even if this group is the additive group of a finite field or a finite ring, we are still only interested in the code being closed under the additive operation.     Throughout the paper, we assume that  the ring $R$ is  a finite Frobenius ring.  We assume also that $R$ is commutative. 
If the ring $R$ is also a field, then linear codes are vector spaces over the finite field; see \cite{Huffman} for a description of codes over fields and \cite{mybook} for a description of codes over rings. In this paper, even though we consider codes over finite fields, in order to apply the different statements, we may just consider them over the underlying group $G$ for additive codes, or the underlying ring $R$ for linear codes.

The MacWilliams relations relate the weight enumerator of a code with the weight enumerator of its orthogonal. Therefore, we need to define the orthogonality for both additive and linear codes.

For linear codes over rings, we have the standard Euclidean inner-product. Namely, for $\vv=(v_1,\dots,v_n),$ $\vw=(w_1,\dots,w_n) \in R^n$, where $R$ is a ring, we define
\begin{equation}\label{Eucinnerproduct} [\vv,\vw] = \sum_{i=1}^n v_i w_i. \end{equation}
We say that $\vv$ and $\vw$ are orthogonal if 
$[\vv,\vw]=0$.

For additive codes over groups, we require some additional definitions in order to define the orthogonality. Let $G$ be a finite abelian group. A character of the group $G$ is a homomorphism from the group $G$ to
the multiplicative group of complex numbers, $\CC^*$. Sometimes, characters are defined
as a homomorphism from $G$ to $\QQ/\ZZ$.  However,  we shall use complex numbers as the image to maintain uniformity with the work done in \cite{mybook}, \cite{bigduality}, \cite{Involve}, \cite{doughertyeven}, \cite{NariLee} and others. The set of characters of $G$ is denoted by $\widehat{G}$.  Namely,
\begin{equation} 
 \widehat{G} = \{ \pi \ | \ \pi \  \text{is a character of } G \}.
 \end{equation}
It is well known that  $\widehat{G}$ 
is a group isomorphic to the group $G$ and that there is no canonical isomorphism between the two groups.  We use this to
construct various dualities that give various orthogonals associated with them.
For a given isomorphism between the group $G$ and its character group $\widehat{G}$, we construct a character table in the following manner.  
Let $\gamma_1,\gamma_2,\dots,\gamma_s$ denote the elements of $G$ and consider the isomorphism $\phi: G \rightarrow \widehat{G}$.  Denote by $\chi_{\gamma_i}$
the image of the element $\gamma_i$ under the isomorphism $\phi$, in other words $\chi_{\gamma_i} = \phi(\gamma_i).$ We then index the rows by $\phi(\gamma_i)$ and the columns by the elements of the group, where the element of the table corresponding to $(\chi_{\gamma_i},\gamma_j)$ is $\chi_{\gamma_i}(\gamma_j).$ 
	
Given an isomorphism of $G$ and $\widehat{G}$ and its corresponding character table, we refer to this as a duality of $G^n$. Since there are different isomorphisms between the groups $G$ and $\widehat{G}$, there are different characters and therefore different dualities.	
 Notationally, we say that $M$ is a duality when $M$ is the character table corresponding to an isomorphism between $G$ and $\widehat{G}$. For $\vv=(v_1,\dots,v_n),\vw=(w_1,\dots,w_n)\in G^n$ we define
\begin{equation}\label{Minnerproduct} [\vv,\vw]_M = \prod_{i=1}^{n}\chi_{v_i}(w_i). \end{equation}
 
Note that, unlike the inner product defined in Equation \ref{Eucinnerproduct}, $[\vv,\vw]_M$ is not an element of $G$ but rather a complex number with norm 1. We say that $\vv,\vw\in G^n$ are orthogonal under the duality $M$ if $[\vv,\vw]_M =1$.

Note that when considering the orthogonality of elements in linear codes over a ring $R$, we always apply the Euclidean inner-product given in Equation \ref{Eucinnerproduct}. However, when considering the orthogonality of elements of additive codes, we need to fix a duality $M$.   We can now define the orthogonal for both additive and linear codes.  

\begin{definition}
Let $G$ be a finite abelian group, and let $R$ be a finite Frobenius ring. 
\begin{itemize}

\item  Fix a duality $M$ of $G$, that is, an isomorphism of $G$ and $\widehat{G}.$ If $C$ is a code over $G$ of length $n$, then
\begin{equation} \label{CM} C^M = \{ \vv\in G^n \, | \, [\vv,\vw]_M =
1, \forall   \vw \in C \}. \end{equation} 

\item  If $C$ is a code over $R$ of length $n$, then
\begin{equation} \label{Cperp} 
C^\perp = \{ \vv \in R^n \ | \ [\vw,\vv]=0, \forall \vw \in C\}.
\end{equation} 
\end{itemize} 
We denote by $C^*$ the orthogonal of $C$; that is, $C^*=C^M$, if $C$ is a code over $G$ or $C^*=C^\perp$, if $C$ is a code over $R$.
\end{definition}  

It is well known (see \cite{mybook}) that $C^M$ is an additive code regardless of whether $C$ is additive. Similarly, $C^\perp$ is a linear code regardless of whether $C$ is linear.  
\medskip

Next, we need to define the complete weight enumerator and the Hamming weight enumerator.  Let $A$ be an alphabet that contains an additive identity $0$. For a given vector $\vc=(c_1,\dots,c_n)\in A^n$, the Hamming weight of $\vc$ is 
$wt(\vc) = | \{i \in\{1,\dots,n\}\ | \ c_i \neq 0 \}|$. %, where $0$ is the identity of the group.   
That is,  the Hamming weight of a vector is its number of non-additive-identity elements. The Hamming distance is defined as $d(\vc_1,\vc_2)=w(\vc_1-\vc_2)$ for any $\vc_1,\vc_2\in R^n$. 

\begin{definition} \label{we}
Let $A$ be any alphabet of ordered elements, $A=\{a_0,\dots,a_r\}.$ For $a\in A$, define $\iota(a)=j$ if $a=a_j$. Let $C$ be a code over $A$ of length $n$. 
\begin{itemize} 
\item  The complete weight enumerator of $C$ is defined as 
%For a code $C$ in $A^n$, 
\begin{equation}\label{eq:cwe}
cwe_C(x_0,x_1,\dots,x_r) = \sum_{(c_1,\dots,c_n) \in C} \prod_{i=1}^n  x_{\iota(c_i)}.
\end{equation}
\item 
The Hamming weight enumerator of $C$ is defined as 
\begin{equation}\label{eq:W_C}
W_C(x,y) = \sum_{\vc \in C} x^{n-wt(\vc)}y^{wt(\vc)}.
\end{equation}
  \end{itemize} 
\end{definition} 

It is immediate that
$$W_C(x,y) = cwe_C(x,y,y,\dots,y).$$

Note that $W_C(1,1)=|C|$.

\begin{example}
Let $A=\FF_4=\{0,1,w,1+w\}$ and $C=\{(0,0),(1,w),(w,w),(1+w,0)\}$. Then, we have
\begin{align*}
cwe_C(x_0,x_1,x_2,x_3)&=x_{\iota(0)}x_{\iota(0)}+x_{\iota(1)}x_{\iota(w)}+x_{\iota(w)}x_{\iota(w)}+x_{\iota(1+w)}x_{\iota(0)}\\&=x_0^2+x_1x_2+x_2^2+x_0x_3, \\ 
& \textnormal{ and }&\\
   W_C(x,y)&=x^2+2y^2+xy=cwe(x,y,y,\dots,y).
\end{align*}
\end{example}

We need to give a characterization of the Frobenius rings necessary for the MacWilliams relations. Let $R$ be a finite commutative Frobenius ring.  Let $\phi :R \rightarrow \widehat{R}$ be a module isomorphism.
Set $\chi = \phi(1)$ so that $\phi(r) = \chi_r$ for $r \in R.$  We call this character $\chi$ a generating character for $\widehat{R}.$  
In this case, for $a,b \in R$ we have $\chi_a(b) = \chi(ab)$.
We have the following immediate consequence.
\begin{proposition}\cite{Wood}
The finite commutative ring $R$ is Frobenius if and only if $\widehat{R}$ has a generating character.
\end{proposition}

For a vector $(x_0,x_1,\dots,x_n)\in R^{n+1}$ and a matrix $M$ being a character table, we denote by $M \cdot (x_0,x_1,\dots,x_{r})$ the operation $(M \cdot (x_0,x_1,\dots,x_{r})^t)^t$. Now we see that the action of the matrix $M$ on the weight enumerator gives the MacWilliams relation. We state the MacWilliams relations for additive and linear codes. The one for linear codes was given in \cite{Wood} and for additive codes in \cite{bigduality}.  

% \begin{theorem}{\bf OLD}   \label{theorem:GroupMacWilliams} 
% Let $G$ be a finite abelian group, $|G|=r+1$, with duality $M$, and let $R$ be a finite commutative Frobenius ring, $|R|=r+1$, with generating character $\chi$.
% \begin{itemize}
% \item  Let $C$ be an additive code over $G$. Then, the complete weight enumerator and the Hamming weight enumerator of its
% orthogonal $C^M$ are given by
% \begin{equation} cwe_{C^M} (x_0,x_1,\dots,x_{r})= \frac{1}{|C|}cwe_C(M \cdot (x_0,x_1,\dots,x_{r})), \end{equation} 
% and 
% \begin{equation}
% W_{C^M}(x,y) = \frac{1}{|C|} W_C(x+ry,x-y),
% \end{equation} 
% respectively.
% \item Let $C$ be a linear code over $R$. Then, the complete weight enumerator and the Hamming weight enumerator of its
% orthogonal $C^\perp$ are given by
% \begin{equation} \label{CCWE}  cwe_{C^\perp} (x_0,x_1,\dots,x_{r})= \frac{1}{|C|}cwe_C(T \cdot (x_0,x_1,\dots,x_{r})), \end{equation} 
% where $T_{a,b} = \chi(ab),$ 
% and 
% \begin{equation}\label{eq:WC-perp}
% W_{C^\perp}(x,y) = \frac{1}{|C|} W_C(x+ry,x-y),
% \end{equation} 
% respectively.
% \end{itemize} 
% \end{theorem}

\begin{theorem}  \label{theorem:GroupMacWilliams} 
Let $A=\{a_0,\dots,a_r\}$ be an alphabet and $C$ a code over $A$ of length $n$. Let $M$ be a given duality $M_{a_i,a_j}=\chi_{a_i}(a_j)$ if $A=G$ a finite abelian group, and $M_{a_i,a_j}=\chi(a_ia_j)$ if $A=R$ a finite commutative Frobenius ring with generating character $\chi$. Then, the complete weight enumerator and the Hamming weight enumerator of the
orthogonal code $C^*$ of $C$ are given by
\begin{equation} \label{CCWE} cwe_{C^*} (x_0,x_1,\dots,x_{r})= \frac{1}{|C|}cwe_C(M \cdot (x_0,x_1,\dots,x_{r})), \end{equation} 
and 
\begin{equation}\label{eq:C*}
W_{C^*}(x,y) = \frac{1}{|C|} W_C(x+ry,x-y),
\end{equation} 
respectively, where $C^*=C^M$ if $A=G$ and $C^*=C^\perp$ if $A=R$.
\end{theorem}

An important consequence of the MacWilliams relations, found by setting the variables equal to $1$ in Equations \ref{eq:cwe} and \ref{eq:W_C}, is that $|C||C^*|=|A|^n$. That is,  $|C||C^M| = |G|^n$ for additive codes, and $|C||C^\perp| = |R|^n$ for linear codes.            

\subsection{Linear codes over a chain ring and $R$-additive codes over a field}\label{section:chain}

In the context of codes over rings, the particular case of codes over finite chain rings is especially relevant. See \cite{mybook} for details for foundational results about codes over chain rings. Let $R$ be a finite commutative ring with identity $1\not =0$. We say that $R$ is a chain ring of index $e$ if there exists $\gamma \in R$ such that the maximal ideal ${\mathfrak m} = \langle \gamma \rangle $
and all of the ideals $\langle \gamma^i \rangle$, $1\leq i\leq e$, are in the chain 
$$ \{ 0 \} \subseteq \langle \gamma^{e-1} \rangle \subseteq \langle \gamma^{e-2} \rangle 	 
	\subseteq \cdots \subseteq \langle \gamma^{2 } \rangle \subseteq \langle \gamma \rangle \subseteq R.$$ 
	Denote the ideal $\langle\gamma^{e-i}\rangle$ by $\fa_i$. Then we have
	$$\{ 0 \} = \fa_0 \subseteq \fa_1 \subseteq \fa_2 \subseteq \cdots \subseteq \fa_{e-1} \subseteq \fa_e =R.$$

We have that $R/\langle \gamma \rangle$ is a isomorphic to a finite field $\mathbb{F}_{q}$ for some $q=p^{m}$, $p$ prime. For $r\in R/\langle \gamma \rangle$, denote by $\bar{r}\in\F_q$ its image under this isomorphism. Let $T=\{e_{0},\dots,e_{q-1}\}\subseteq R$ such that $e_0=0$, $e_1=1$ and $\overline{e_{i}}\neq\overline{e_{j}}$ for all $i,j \in \{0,1,\dots,q-1\}$, $i\neq j$. Any element $r\in R$ can be written as $r=\sum_{i=0}^{e-1}{r}_i\gamma^i$, where ${r}_i\in T$ for $i\in \{0,\dots,s-1\}$. We refer to $[\bar{r}_{0},\dots,\bar{r}_{s-1}]_{\gamma}$ as the $\gamma$-adic representation of $r$.

\begin{lemma}[\cite{MacD}]
    For any $a\in R$, there is a unique integer $i$, $0 \leq i\leq e$, such that $a =\mu\gamma^i$ with $\mu$ a unit. The unit $\mu$ is unique modulo $\gamma^{e-i}$.
\end{lemma}

\begin{corollary}[\cite{SalageanChain}]
If $1\leq i< j\leq e$, and $\gamma^ic\in\langle \gamma^j\rangle$, then $c\in\langle \gamma^{j-i}\rangle$. In particular, if $\gamma^ic=0$, then $c\in\langle \gamma^{e-i}\rangle$.    
\end{corollary}

Let $R$ be any ring and $a\in R$. The order of $a$, denoted by $\operatorname{ord}(a)$ is the smallest natural $c$ such that $c\cdot a=0$. For $R$ a finite chain ring and $a\in R$, we define the order in the chain ring of $a$ as $\gamma^i$, where $0 \leq i\leq e$ such that $a =\mu\gamma^{e-i}$ with $\mu$ as a unit. If $a\in R$ is of order $\gamma^i$, then $\gamma^ia=0$. We denote this order by $\operatorname{ord}_R(a)$. Note that when $R=\Z_{p^s}$, where $p$ is a prime number, $\operatorname{ord}(a)=\operatorname{ord}_R(a)$. In this paper, we will refer $\operatorname{ord_R(a)}$ simply as the order of $a$.

\medskip

Let $\fb_i = \fa_i \setminus \fa_{i-1} = \{  r \in \fa_i \ | \ r \notin \fa_{i-1} \}.$ 
We note that $\fb_0,\fb_1,\fb_2,\dots,\fb_e$ form a partition of the ring $R$.  That is $\cup_{i=0}^e \fb_i =R$ and $\fb_i \cap \fb_j = \emptyset $ if $i\neq j.$   We also note that $\fb_e$ consists of all units in $R$.

\begin{remark}\label{remark:OrderExact_i}
    %Since $\fa_i=\langle \gamma^{e-i}\rangle$,
    The elements in $\fa_i$ are all the elements of $R$ of order at most $\gamma^i$. 
    %Since $\fb_i = \fa_i \setminus \fa_{i-1}$,
    The elements of $\fb_i$ are the elements of $\fa_i$ of order exactly $\gamma^i$.
\end{remark}

As a corollary of the remark, and the definition of the sets $\fa_i,\fb_i$, we have the following statement.
\begin{lemma}\label{lemma:fafb}
 Let $a\in R$ and $i$ such that $1\leq i\leq e$.
 \begin{itemize}
 \item If $a\in\fa_i$, then $a=c\gamma^{e-i}$, for $c\in R$.
  \item If $a\in\fb_i$, then $a=\mu\gamma^{e-i}$, for $\mu$ a unit.
 \end{itemize}
\end{lemma}

As shown in \cite{SalageanChain}, a linear code $C$ over a chain ring $R$ is equivalent to a code with a generator matrix of the following form:
\begin{equation}
\label{generator}
\left(  \begin{array}{ccccccc}
 I_{k_0} &  A_{0,1}  & A_{0,2} &  A_{0,3}  & \cdots& \cdots & A_{0,e}  \\
0       & \gamma I_{k_1} & \gamma A_{1,2}& \gamma A_{1,3} & \cdots& \cdots &
\gamma A_{1,e} \\
0       & 0        & \gamma^2 I_{k_2}& \gamma^2 A_{2,3} & \cdots& \cdots & \gamma^2 A_{2,e} \\
\vdots  & \vdots   & 0       & \ddots   & \ddots&        & \vdots     \\
\vdots  & \vdots   & \vdots  & \ddots   & \ddots& \ddots & \vdots     \\
0       & 0        & 0       & \cdots   & 0      & \gamma^{e-1} I_{k_{e-1}} &
\gamma^{e-1} A_{e-1,e}
\end{array} \right).
\end{equation}

Without loss of generality, we assume that the code $C$ has a generator matrix of this form. 
We say that the code $C$ has type $(k_0,k_1,\dots,k_{e-1}).$ It follows immediately that the number of elements in 
 a  code  with this generator matrix is
\begin{equation}\label{anasize}
|C| =|R/{\mathfrak m}|^{\sum_{i=0}^{e-1} (e-i)k_i}=
|\FF_q|^{\sum_{i=0}^{e-1} (e-i)k_i} =q^{\sum_{i=0}^{e-1} (e-i)k_i} .
\end{equation}
Now we relate the type of a code over a chain ring to the type of its orthogonal code. We define $k_e= n -\sum_{i=0}^{e-1} k_i$.

\begin{theorem} \label{anyortho}
Let $C$ be a linear code of length $n$ over the chain ring $R$ of type $(k_0,k_1,\dots,$ $k_{e-1})$, and let $M$ be a duality on the additive group of the chain ring $R$.  The orthogonal code $C^M$ is a code of type
 $(k_e,k_{e-1},k_{e-2},\dots,k_{1}).$ 
 \end{theorem} 
 \begin{proof}
 If $C$ is linear then $C^\perp = C^M$ for any duality $M$ (see \cite{bigduality}) and the result for $C^\perp$ is well known (see \cite{SalageanChain}). 
 \end{proof}

We now consider the order of an element in $R^n$ and, given a code over $R$ and $1\leq i\leq e$, we consider the sets of codewords of order $\gamma^i$, and the set of codewords of order at most $\gamma^i$. The following lemma is clear by the definition of the order of a coordinate.

\begin{lemma} \label{orders} 
Let $\vv = (v_1,v_2,\dots,v_n)\in R^n.$ Then the order of the vector $\vv$ is 
$$\operatorname{ord}_R(\vv)=\max \{\operatorname{ord}_R(v_i)\}.$$
\end{lemma}
% \begin{proof}
% Let $c$ be a natural number and let $c x = \sum_{i=1}^c x.$
% If $c \vv = c (v_1,v_2,\dots,v_n) = {\bf 0}$ then $c v_i = 0$ for all $i$.  Therefore, the smallest $c$ with $c\vv = {\bf 0}$ is the maximum of all of the orders of the elements.
% \end{proof} 

Let $C$ be a linear code over $R$ of length $n$. Let $G$ be a generator matrix of $C$ as in Equation \ref{generator}, and let $\fB_0= \fA_0= \{ {\bf 0} \}.$ Let $\fA_i$
be the set of vectors generated by 
{\tiny \begin{equation} \label{genB}  \left(  \begin{array}{cccccccc}
\gamma^{e-i} I_{k_0} & \gamma^{e-i} A_{0,1}  & \gamma^{e-i} A_{0,2} & \gamma^{e-i} A_{0,3}  & \cdots& \cdots & \cdots & \gamma^{e-i} A_{0,e}  \\
0       & \gamma^{e-i} I_{k_1} & \gamma^{e-i} A_{1,2}& \gamma^{e-i} A_{1,3} & \cdots &\cdots \cdots & \cdots&
\gamma^{e-i} A_{1,e} \\
0       & 0        & \gamma^{e-i} I_{k_2}& \gamma^{e-i} A_{2,3} & \cdots& \cdots & \cdots &\gamma^{e-i} A_{2,e} \\
\vdots  & \vdots   & 0       & \ddots   & \ddots &     \ddots &  \ddots & \vdots     \\
\vdots  & \vdots   & \vdots  & \ddots   & \ddots& \ddots &  \ddots & \vdots     \\
0       & 0        & 0       & \gamma^{e-i} I_{k_{e-i}}  & \cdots  & \cdots     \cdots & \cdots   &
\gamma^{e-i} A_{e-i,e} \\
0       & 0        & 0       & \cdots &  \gamma^{e-{i+1}} I_{k_{e-{i+1} }}   & \cdots &\cdots       &
\gamma^{e-{i+1}} A_{e-i+1 ,e} \\
\vdots  & \vdots   & \vdots  & \ddots   & \ddots& \ddots & \vdots    & \vdots  \\
0       & 0        & 0       & \cdots   & \cdots & 0      & \gamma^{e-1} I_{k_{e-1}} &
\gamma^{e-1} A_{e-1,e}
\end{array} \right), 
\end{equation}}
and let $\fB_i=\fA_i\setminus \fA_{i-1}.$ %that are not in $\fB_{s}, s < i.$
It follows immediately that if $j \neq k$, $\fB_j \bigcap \fB_k = \emptyset $ and $\bigcup_{i=0}^e \fB_i =C$. That is, $\fB_0,\fB_1,\fB_2,\dots,\fB_e$ form a partition of the code $C$. Moreover, $\fA_i=\cup_{j=0}^{i}\fB_j$ and hence
\begin{equation}\label{eq:BiAi}
\fB_i=\fA_i\setminus(\cup_{j=0}^{i-1}\fB_j).
\end{equation}

By the construction of the subcodes $\fA_i$ and the sets $\fB_i$ we have the following statement.

\begin{lemma}
 Let $C$ be a code over $R$ of length $n$, $\vv\in C$, and $i$ such that $1\leq i\leq e$.
 \begin{itemize}
 \item If $\vv\in\fA_i$, then $\vv=\gamma^{e-i}\vc$, for $\vc\in R^n$.
  \item If $\vv\in\fB_i$, then $\vv=\gamma^{e-i}\vu$, for $\vu\in R^n$ with $\operatorname{ord}_R(\vu)=1$.
 \end{itemize}
\end{lemma}

By the definition of the generator matrix Equation \ref{genB} of $\fA_i$ and the definition of the type, we have the following statement.

\begin{lemma}\label{lemma:typefA}
 Let $C$ be a linear code over $R$ of length $n$, $1\leq i \leq e$, and $\fA_i$ be the subcode of $C$ with generator matrix as in Equation \ref{genB}. Then $\fA_i$ is of type 
 $$
 (0,\stackrel{e-i}\cdots,0,k_0+\dots+k_{e-i},k_{e-i+1},\dots,k_{e-1}).
 $$
\end{lemma}
\begin{example} \label{firstex}
We consider a simple example to illustrate these sets. 
Consider $R$ the finite chain ring $\ZZ_{16}$, with ${\mathfrak m}=\langle2\rangle$ and $|R/{\mathfrak m}|=|\FF_2|$, so $q=2$. Consider the code $C$ generated by the matrix 
\begin{equation} 
\left(  \begin{array}{cccc}
1 & 0 & 0 & 0 \\
0 & 2 & 0 & 0 \\
0 & 0 & 4 & 0 \\
0 & 0 & 0 & 8 \\
\end{array} \right).
\end{equation}
The code has type $(4;1,1,1,1)$ and has 
$2^{4\cdot 1+3\cdot 1+2\cdot 1+1\cdot 1} = 1024$  elements. 
The code $\fB_0 =\fA_0= \{ (0,0,0,0) \}.$
Then $\fA_1$ is the code generated by 
\begin{equation} 
\left(  \begin{array}{cccc}
8 & 0 & 0 & 0 \\
0 & 8 & 0 & 0 \\
0 & 0 & 8 & 0 \\
0 & 0 & 0 & 8 \\
\end{array} \right),
\end{equation}
has type $(4;0,0,0,4)$, and $\fB_1=\fA_1\setminus\fA_0=\fA_1\setminus\fB_0.$
Then $\fA_2$ is the code generated by 
\begin{equation} 
\left(  \begin{array}{cccc}
4 & 0 & 0 & 0 \\
0 & 4 & 0 & 0 \\
0 & 0 & 4 & 0 \\
0 & 0 & 0 & 8 \\
\end{array} \right),
\end{equation}
has type $(4;0,0,3,1)$ and $\fB_2=\fA_2 \setminus \fA_1=\fA_2 \setminus( \fB_1 \bigcup \fB_0).$
Then $\fA_3$ is the code  generated by 
\begin{equation} 
\left(  \begin{array}{cccc}
2 & 0 & 0 & 0 \\
0 & 2 & 0 & 0 \\
0 & 0 & 4 & 0 \\
0 & 0 & 0 & 8 \\
\end{array} \right),
\end{equation}
has type $(4;0,2,1,1)$ and $\fB_3=\fA_3 \setminus \fA_2 =\fA_3 \setminus(  \fB_2 \bigcup  \fB_1 \bigcup \fB_0).$
Finally,  $\fB_4$ is the code  generated by 
\begin{equation} 
\left(  \begin{array}{cccc}
1 & 0 & 0 & 0 \\
0 & 2 & 0 & 0 \\
0 & 0 & 4 & 0 \\
0 & 0 & 0 & 8 \\
\end{array} \right),
\end{equation}
has type $(4;1,1,1,1)$ and $\fB_4=\fA_4 \setminus \fA_3 =\fA_4\setminus(\fB_3 \bigcup  \fB_2 \bigcup  \fB_1 \bigcup \fB_0).$

\end{example}

\begin{example} \label{nextex} 
   Recall that a free code is a code that has a generator matrix of the form $(I_k \ | \ A).$
   Let $C$ be the code generated by this matrix over $\ZZ_{p^e}. $
   Then $\fB= \{ {\bf 0 } \}$ and
   $\fB_i$ consists of the vectors  in  $\langle \left( p^{e-i} I_k \ | \  p^{e-1} A \right) \rangle\setminus \bigcup_{j=0}^{i-1} \fB_i.$
\end{example}

\begin{theorem}\label{counter} 
Let $C$ be a code of length $n$ over a finite chain ring $R$ with maximal ideal $\fm= \langle \gamma \rangle$, where $\gamma$ has index $e$, and $|R/{\mathfrak m}|=|\FF_q|$.
We have that $ | \fB_0|=1$ and
$$
 | \fB_i|  = q^{\sum_{j=0}^{e-i} i k_j + \sum_{j=e-i+1}^{e-1} (e-j)  k_j} - \sum_{j=0}^{i-1}|\fB_j|
$$
for $i\geq 1$.
% \begin{eqnarray*}
%     | \fB_0| &=& 1 \\
%     | \fB_1|  &=& |R/\fm|^{\sum_{i=0}^{e-1}  k_i } -1 \\
       
%                 | \fB_3|  &=& |R/\fm|^{\sum_{i=0}^{e-3}  3  (e-3) k_i}
%                 |R/\fm|^{\sum_{i=e-2}^{e-1} (e-i)  k_i} -|\fB_2|-|\fB_1|- 1 \\
%                 \vdots \\
%                         | \fB_{e-1}|  &=& |R/\fm|^{\sum_{i=0}^{1} (e-1) k_i} |R/\fm|^{\sum_{i=2}^{e-1} (e-i)  k_i}   - |\fB_{e-2} | - |\fB_{e-3} - \cdots |- |\fB_1|  -1 \\
%                         | \fB_{e}|  &=& |R/\fm|^{\sum_{i=0}^{e-1} (e-i) k_i} - |\fB_{e-1} | - |\fB_{e-2} | - |\fB_{e-3} - \cdots |- |\fB_1|  -1
% \end{eqnarray*}
\end{theorem} 
\begin{proof}
    By Equation \ref{eq:BiAi} and, considering that $\fB_j \bigcap \fB_k = \emptyset$ for $j\not=k$, we have that $| \fB_i|  = |\fA_i| - \sum_{j=0}^{i-1}|\fB_j|$. Since $\fA_i$ is of type $
 (0,\stackrel{e-i}\cdots,0,k_0+\dots+k_{e-i},k_{e-i+1},\dots,k_{e-1})$ by Lemma \ref{lemma:typefA}, we obtain the result by Equation \ref{anasize}.
\end{proof}

\begin{example} \label{secondex}
We continue with the code $C$ in Example~\ref{firstex} of type $(4;1,1,1,1)$. We have
\begin{eqnarray*}
    |\fB_0|  &=& 1 \\   
    |\fB_1|  &=& 16 - 1 =15 \\  
    |\fB_2|  &=& 128-15 -1 =112 \\  
    |\fB_3|  &=& 512 - 112 -15 - 1 = 384 \\  
    |\fB_4|  &=& 1024-384-112-15 -1 = 512. \\  
\end{eqnarray*}
We note that $1+15+112 + 384 + 512 = 1024 = |C|.$
\end{example} 
\begin{example} \label{secondnextex}
We continue with Example~\ref{nextex}. Note that the code is of type $(k,0,\cdots,0)$. We have that $|\fA_i|=p^{ik}$, and
\begin{eqnarray*}
    |\fB_0|  &=& 1 \\   
    |\fB_1|  &=& p^k - 1 \\  
    |\fB_2|  &=& p^{2k} - (p^k - 1) - 1 = p^{2k}-p^k  \\  
    |\fB_3|  &=& p^{3k} - (p^{2k} - p^k) - p^k = p^{3k}-p^{2k}  \\
    &\vdots&  \\
    |\fB_e|  &=& p^{ek} - p^{(e-1)k}  \\  
\end{eqnarray*}
Note that $\sum_{j=0}^e |\fB_j| = p^{ek}.$ 
\end{example} 

\begin{theorem} Let $R$ be a finite commutative chain ring with maximal ideal $\fm$ where $|R / \fm | =q.$  Then $|\fb_i|= q^i-q^{i-1}$ for $i>0$ and $|\fb_0|=1.$
\end{theorem}
\begin{proof}
    We can consider the ring $R$ as a code of length $1$ over $R.$  Then $\fB_i = \fb_i.$  Then $\fb_0 = \{ 0 \}$ and so $|\fb_i| =1.$  
    Then $|\fb_1| = q^1-1.$ This gives $|\fb_2| = q^2 - (q-1) -1 = q^2-q.$ Therefore, the statement is true for $i=1,2.$
    Then \begin{eqnarray*} |\fb_i| &=&  q^i - |\fb_{i-1}| - |\fb_{i-2}| - |\fb_{i-3} | - \cdots - |\fb_1| - |\fb_0|  \\
    &=& q^i - (q^{i-1} - q^{i-2}) - (q^{i-2} - q^{i-3}) - \cdots - (q-1) -1    \\
    &=& q^i - q^{i-1}.
        \end{eqnarray*} 
        This gives the result by induction.   
\end{proof}

\begin{theorem} \label{autre} 
    Let $R$ be a finite chain ring with maximal ideal $\fm = \langle \gamma \rangle $ of index $e$  and let $C$ be a code of length $n$ over $R$.  Then $\fB_i$ consists of all elements in the code $C$ of order $\gamma^i$ for $0\leq i\leq e$.% where $t_i$ is the order of $\gamma^{e-i}$. 
\end{theorem}

 \begin{proof}It is clear for $i=0$, since $\fB_0=\{\zero\}$. Let $i$ be an integer,  $1\leq i\leq e$. By the construction of the generator matrix given in Equation~\ref{genB}, $\fA_i$ is the subcode of $C$ containing all the codewords of order at most $\gamma^i$. Since $\fB_i=\fA_i\setminus \fA_{i-1}$, $\fB_i$ contains all the codewords of order exactly $\gamma^i$.
 \end{proof}

 \begin{corollary} 
 Let $R$ be a finite chain ring with maximal ideal $\fm = \langle \gamma \rangle $ of index $e$  and let $C$ be a code of length $n$ over $R$. 
     The code $\fB_e$ consists of all vectors in $C$ that have a unit in a coordinate.
 \end{corollary}

 \begin{proof}
Theorem~\ref{autre} gives that 
the codewords in $\fB_e$ are those codewords of order $e$. By Lemma \ref{orders}, those codewords must have a coordinate of order $e$; that is, a coordinate of the form $\mu\gamma^{e-e}=\mu$, for $\mu$ a unit.      \end{proof}

\begin{example} \label{secondex2}
We continue with Example~\ref{firstex}. We have
that $\fB_0$ consists of only the zero vector which has order $1.$ The code $\fB_1 $ has vectors of order $2$, the code $\fB_2 $ has vectors of order $4$, 
the code $\fB_3 $ has vectors of order $8$, and the code $\fB_4 $ has vectors of order $16$.  We note that $\fB_4$ has cardinality $512$ which is all the elements generated by
$\lambda_0(1,0,0,0) + \lambda_1 (0,2,0,0) + \lambda_2(0,0,4,0) + \lambda_3 (0,0,0,8)$, where $\lambda_0 \in \{1,3,5,7,9,11,13,15\}$.  There are $8(8)(4)(2) = 512 $ such vectors.
\end{example}

\begin{example} \label{secondnextex3}
We continue with Example~\ref{nextex}. We have
that $\fB_0$ consists of only the zero vector which has order $1.$ The code $\fB_1 $ has vectors of order $p$, the code $\fB_2 $ has vectors of order $p^2$, 
the code $\fB_3 $ has vectors of order $p^3$, and the code $\fB_e $ has vectors of order $p^e$.
\end{example} 

Finally, we consider the homogeneous weight of codewords in a code over a chain ring $R$, and we relate this weight to the Hamming weight of its image under a generalized Gray map. We define also such a generalizations.

\medskip

Let $C$ be a code over a chain ring $R$ and let $\vc=(c_1,\dots,c_n)\in C$. We can consider the Hamming weight of $\vc$, $w(\vc)$, previously defined, and also the homogeneous weight of $\vc$, $w_{Hom}(c)=\sum_{i=0}^n w_{hom}(c_i)$, where
$$w_{hom}(c_i)
=\begin{cases} 0 & \text{if $c=0$,} \\
  (q-1)q^{e-2} & \text{if $c\neq0$ and $c\notin\langle \gamma^{e-1}\rangle$,}\\
  q^{e-1} &\text{if $c\neq0$ and $c\in\langle \gamma^{e-1}\rangle$.}
\end{cases}$$
The homogeneous distance is defined as $d_{Hom}(\vc_1,\vc_2)=w_{Hom}(\vc_1-\vc_2)$ for any $\vc_1,\vc_2\in R^n$.

A Gray map for any finite chain ring $R$ is $\phi:R\longmapsto \mathbb{F}_{q}^{q^{e-1}}$ such that
\begin{gather}\label{eq:GrayMapCarlet}
\phi(r)=(\bar{r}_0,\bar{r}_1, \dots, \bar{r}_{e-1})\begin{pmatrix} Y \\ \one \end{pmatrix},
\end{gather}
where  $[\bar{r}_{0},\dots, \bar{r}_{s-1}]_\gamma$ is the $\gamma$-adic representation of $r$, and $Y$ is a matrix of size $(e-1)\times q^{e-1}$ whose columns are all the vectors in $\F_{{q}}^{e-1}$ (see \cite{GrayIsometry}). We define $\Phi :R^n \rightarrow \F_{q}^{nq^{(e-1)}}$ as the component-wise extended map of $\phi$. This generalization is given with a different definition in \cite{Jitman} where it is proved that $\Phi$ is an isometry from $(R^n, d_{Hom})$ to $(\F_{q}^{nq^{(e-1)}}, d)$. Therefore, we have that for $\vc\in R^n$
\begin{equation}\label{eq:phi-isometry}
w(\Phi(\vc))=w_{Hom}(\vc).
\end{equation}

Let $C$ be a linear code of length $n$ over a chain ring $R$. Then, the code $\Phi(C)$ is called an $R$-additive code. Note that $\Phi(C)$ is a code over $\FF_q$ of length $nq^{(e-1)}$ which is not necessarily linear.

\section{Weight enumerators with different equivalence relations}  \label{tre}

In this section, we state and prove our main tool in this paper; namely the weight enumerator of a code with respect to a given equivalence relation. In the different subsections, we explore the weight enumerator obtained for various equivalence relations. Section \ref{subsection:Hamming} gives the weight enumerator related to the  Hamming weight  for codes over any alphabet $A$. Section \ref{subsection:Symmetric} gives the symmetric weight enumerator for codes over a group $G$ and Section \ref{subsection:lambda} is a generalization of the symmetric weight enumerator, and it is given for codes over commutative Frobenius rings $R$. Finally, in Section \ref{subsection:CRW} we study some weight enumerators for codes over a chain ring $R$,  both of which are related to the order of elements; either the order of coordinates or the order of codewords. In the first case, the obtained weight enumerator of a code $C$ is related to the Hamming weight enumerator of the code $\Phi(C)$.

  Let $A=\{a_0,\dots,a_r\}$ be an alphabet with an  equivalence relation $\equiv$, and equivalence classes $A_1,A_2,\dots,A_s$; that is, subsets of $A$ formed by grouping all elements that are equivalent to each other under $\equiv$.

  Let $A=A_1\cup A_2 \cup \dots\cup A_s$, where $A_1,A_2,\dots,A_s$ are non-empty,  disjoint sets. We say that ${\cal{P}}=\{A_1,A_2,\dots,A_s\}$ is a partition of $A$. Note that, for an equivalence $\equiv$, its equivalence classes give a partition of $A$. On the other hand, if we have a partition ${\cal{P}}=\{A_1,A_2,\dots,A_s\}$, we can define the equivalence relation $\equiv_\cP$, where two elements $a,b\in A$ are equivalent, $a\equiv_\cP b$, if and only if $a,b\in A_j$ for some $j\in\{1,\dots,s\}$.

\begin{definition}
   Let $A=\{a_0,\dots,a_r\}$ be an alphabet with equivalence relation $\equiv$. Let $A_1,A_2,\dots,A_s$ be the equivalence classes. For $\vc=(c_1,\dots,c_n)\in A^n$ and $i\in\{1,\dots,s\}$, denote by $N_i(\vc) = | \{ j\in\{1,\dots,n\} \ | \ c_j \in  A_i \}|.$ Define the equivalence weight enumerator for a code $C$ over $A$ with respect to the equivalence $\equiv$ as
\begin{equation} \label{eq:EW_C}
EW_C(x_1,x_2,\dots,x_s) = \sum_{\vc \in C} \prod_{i=1}^s x_i^{N_i(\vc)}.
\end{equation}   
\end{definition}

In \cite{Heide}, Gleusing-Luerssen defines a dual partition to a partition of a finite abelian group on the character group which we describe  below. 
% and then show how it gives a corollary to our previous lemma. 
Let $A=G$ be a group and ${\cal{P}}=\{A_1, A_2,\dots, A_s\}$ be a partition of $G$. Fix an isomorphism $\phi: G \rightarrow \widehat{G}$, and consider $\chi_{\gamma_i} = \phi(\gamma_i),$ for $\gamma_i\in G$. 
The dual partition $\widehat{\cal{P}}=\{\widehat{B_1}, \widehat{B_2},\dots,\widehat{B_r}\}$ is the partition of $\widehat{G}$ defined by the equivalence relation $\chi_a\equiv_{\widehat{\cal{P}}} \chi_{a'}$ if and only if $\sum_{b \in A_j} \chi_a(b) = \sum_{b \in A_j} \chi_{a'}(b)$ for all $j\in\{1,\dots,s\}$. The partition is called reflexive (or Fourier-reflexive) if $\widehat{\widehat{\cal{P}}}=\cal{P}.$

\begin{lemma}[\cite{Heide}]\label{lemma:reflexive}
    Let $\cP$ be a partition of $A$. Then, $\cP$ is reflexive if and only if $|\cP|=|\wcP|$.
\end{lemma}

\begin{definition} Let $A=G$ be a group with equivalence relation $\equiv$, let ${\cal{P}}=\{A_1, A_2,\dots, A_s\}$ be the partition of the equivalence classes, and $\widehat{\cal{P}}$ be its dual partition with equivalence relation $\equiv_{\widehat{\cal{P}}}$. We say that $\cal{P}$ is autodual if we have that $a\equiv a'$ if and only if $\chi_a\equiv_{\widehat{\cal{P}}} \chi_{a'}$ for all $a,a'\in A$.
\end{definition}

By the last definition, if ${\cal{P}}=\{A_1, A_2,\dots, A_s\}$ is an autodual partition, then
$$
\wcP=\{\{\chi_{a_1}\mid a_1\in A_1\}\}, \dots,\{\chi_{a_s}\mid a_s\in A_s\}\}
$$
\begin{corollary}\label{coro:autoDualReflexive}
Let $\cP$ be a partition of $A$. If $\cP$ is autodual, then $\cP$ is reflexive.
\end{corollary}

\begin{proof}
    By definition, if $\cP$ is autodual, the number of classes of the equivalences $\equiv$ and $\equiv_\wcP$ coincide. Then $\cP$ is reflexive by Lemma~\ref{lemma:reflexive}.
\end{proof}

The following example, given in \cite{Heide}, shows that the converse of Corollary~\ref{coro:autoDualReflexive} is not true in general.

\begin{example}
Let $A=\Z_6=\{0,1,2\}$, let $\omega$ be a primitive complex sixth root of unity, and set $\chi_a(b)=\omega^{ab}$, for $a,b\in\Z_6$. Consider the partition into $\cP=\{A_1,A_2,A_3\}=\{\{0\},\{1,3,5\},\{2,4\}\}$. We have that
\begin{align*}
&\sum_{a\in A_1}\chi_0(a)=1,\;\sum_{a\in A_2}\chi_0(a)=3,\; \sum_{a\in A_3}\chi_0(a)=2,\\
&\sum_{a\in A_1}\chi_b(a)=1,\;\sum_{a\in A_2}\chi_b(a)=0,\; \sum_{a\in A_3}\chi_0(a)=-1,\\
&\sum_{a\in A_1}\chi_0(a)=1,\;\sum_{a\in A_2}\chi_0(a)=-3,\; \sum_{a\in A_3}\chi_0(a)=2,
\end{align*}
and hence $\wcP=\{\{\chi_0\},\{\chi_1,\chi_2,\chi_4,\chi_5\},\{\chi_3\}\}$. By Lemma~\ref{lemma:reflexive}, we have that $\cP$ is reflexive, but it not autodual.
\end{example}

\begin{theorem} \label{theorem:important} 
 Let $A=\{a_0,\dots,a_r\}$ be an alphabet with an equivalence $\equiv$, and $C$ be a code over $A$ of length $n$. 
Let $M$ be the matrix given for additive or linear codes; that is, $M_{a,b} = \chi_a(b)$ for the additive case and $M_{a,b} = \chi(ab)$ for Frobenius rings. 
Let $\cP=\{A_1,A_2,\dots,A_s\}$ be the partition of the equivalence classes. If $\cP$ is autodual, then the MacWilliams relations exist for the weight enumerator $EW$ and 
 \begin{equation}\label{eq:EW-perp} EW_{C^*} (x_1,x_2,\dots,x_{s})= \frac{1}{|C|}EW_C(\overline{M} \cdot (x_1,x_2,\dots,x_{s})), \end{equation} 
 where $\overline{M}$ is an $s\times s$ matrix where $\overline{M}_{A_i,A_j} = \sum_{b \in A_j}  M_{a,b}$, for any element $a$ in $A_i$. 
\end{theorem}

\begin{proof}
Let $i\in\{1,\dots,s\}$. Assume $\cP$ is autodual; that is, $\sum_{b \in A_j} M_{a,b} = \sum_{b \in A_j} M_{a'b}$ whenever $a,a' \in A_i$. Then, the action of the matrix $M$ is the same for all elements of the equivalence classes $A_1,A_2,\dots,A_s.$  Therefore, the MacWilliams relations given by $\overline{M}$ hold.
\end{proof}

In \cite{Heide}, the elements $\overline{M}_{A_i,A_j}$ are called the generalized Krawtchouk coefficients, and $\overline{M}$ is the generalized Krawtchouk matrix of $(\cP,\wcP)$. It is also shown that reflexive partitions give rise to MacWilliams identities for weight enumerators using both the partition $\cal{P}$ and the dual partition $\widehat{\cal{P}}$. This is a different setting than the setting which we have described since we consider only one  partition that is related to a specific equivalence relation.

\subsection{Weight enumerators with Hamming weight relations}\label{subsection:Hamming}

Let $A=\{a_0,\dots,a_r\} $ be an alphabet, where $a_0=0$ is the additive identity. Consider the relation $\sim_H$ defined by $a\sim_H a'$, for $a,a'\in A$, if $wt(a)=wt(a')$. Note that $wt(a_0)=0$, and $wt(a_i)=1$ for all $i\in\{1,\dots,r\}$. Therefore, we have the classes $A_1=\{0\}$ and $A_2=\{a_1,\dots,a_r\}$. We define then the equivalent weight enumerator given in Equation \ref{eq:EW_C} related to the relation $\sim_H$ as
\begin{equation} \label{eq:HW_C}
HW_C(x_1,x_2) = \sum_{\vc \in C} x_1^{N_1(\vc)}x_2^{N_2(\vc)}.
\end{equation}

\begin{example}\label{ex:HammingF3}
Let $A=\FF_3=\{0,1,2\}$, let $\omega$ be a primitive complex third root of unity, and set $\chi_a(b)=\omega^{ab}$, for $a,b\in\FF_3$. Consider the equivalence class that gives the Hamming weight.  Namely, the alphabet has a partition into $A_1 = \{0\}$ and $A_2 = \{ a \ | \ a\neq 0 \}=\{1,2\}.$  
Then, over $\FF_3$ the MacWilliams relations for the complete weight enumerator are given by the matrix 
$$ M = \left( \begin{array}{ccc} 1 & 1 & 1 \\ 1 & \omega & \omega^2 \\ 1 & \omega^2 & \omega \end{array}  \right).$$ 
Now $\sum_{b \in A_2} M_{1,b} = \omega+\omega^2 = -1$ and $\sum_{b \in A_2} M_{2,b} = \omega^2+\omega = -1$
and $\sum_{b \in A_2} M_{0,b} =2$.
This gives that the matrix $\overline{M}$ in this case is $$ 
\overline{M}=\left( \begin{array}{cc} 1 & 2 \\ 1 &  -1 \end{array}  \right),$$ which is the MacWilliams relation for the Hamming weight enumerator over $\FF_3$.  
\end{example}
\begin{example}\label{ex:HammingZ4}
Let $A=\Z_4=\{0,1,2,3\}$, $\sqrt{-1}$, and set $\chi_a(b)=i^{ab}$, for $a,b\in\Z_4$. Consider the equivalence class that gives the Hamming weight.  Namely, the alphabet has a partition into $A_1 = \{0\}$ and $A_2 = \{1,2,3\}.$  Over $\ZZ_4$ the MacWilliams relations are given by the matrix 
$$ M = \left( \begin{array}{cccc} 1 & 1 & 1 & 1  \\ 1 &i & -1  & - i \\ 1 & -1 & 1 & -1  \\
1 &-i & -1  &  i  \\ 
 \end{array}  \right).$$  
Now $\sum_{b \in A_2} M_{1,b} = i-1 -i = -1$, $\sum_{b \in A_2} M_{2,b} = -1+1 - 1 = -1$, $\sum_{b \in A_2} M_{w,b} = -i -1 + i  = -1$,
and $\sum_{b \in A_2} M_{0,b} =3$.
This gives that  $$ 
\overline{M}=\left( \begin{array}{cc} 1 & 3 \\ 1 &  -1 \end{array}  \right),$$ which gives the MacWilliams relation for the Hamming weight enumerator over $\ZZ_4$. \end{example}

We can generalize the results obtained in Examples \ref{ex:HammingF3} and \ref{ex:HammingZ4} in the following statement.
% \begin{proposition}{\bf OLD}
%     Let $A=\{a_0,\dots,a_r\}$ be a group or a ring with $a_0=0$, and $\sim_H$ the equivalent relations given by the Hamming weight of the elements of $A$. Let $\chi_a$ be the character associated with the element $a\in A$ such that the MacWilliams relations are given by the matrix 
%     $$ M = \left( \begin{array}{ccc} 1 & 1 \cdots 1  \\ \vdots &\multirow{2}{*}M'\\ 
% 1 &  \\ 
%  \end{array}  \right).$$
%  Then, %\sum_{i=1}^{r} M'_{i,j}=-1$, for all $j\in\{1,\dots,r\}$, and 
%  (\ref{eq:EW_C}) is satisfied by the matrix 
%    $$   \overline{M} = \left( \begin{array}{cc} 1 & r  \\ 1 & -1  \\ 
%  \end{array}  \right).$$
%  Moreover, 
% \begin{equation}\label{eq:HWC-perp2}
% HW_{C^M}(x_1,x_2) = \frac{1}{|C|} HW_C(x_1+rx_2,x_1-x_2).
% \end{equation} 
% \end{proposition}
% \begin{proof}
%  Since $\chi$ is a character we have $\sum_{b\in A} \chi_a(b) = 0.$
%  Therefore,  $\sum_{i=1}^{r} M'_{i,j}=-1.$ Then Lemma~\ref{important} is satisfied. By appling (\ref{eq:EW-perp}) with matrix $\overline{M}$, we obtain $HW_{C^M}(x_1,x_2) = \frac{1}{|C|} HW_C(x_1+rx_2,x_1-x_2).$
% \end{proof}

\begin{proposition}\label{prop:Hamming1}
    Let $A=\{a_0,\dots,a_r\}$ be a group or a ring with $a_0=0$ and with generating character $\chi$. Let $M$ be the matrix related to $\chi$. Let $A_1,A_2$ be the equivalence classes under the equivalence relation $\sim_H$. 
 Then,  
\begin{equation}\label{eq:HWC-perp}
HW_{C^*}(x_1,x_2) = \frac{1}{|C|} HW_C(x_1+rx_2,x_1-x_2).
\end{equation} 
\end{proposition}
\begin{proof}
 Let $\chi_a$ be the character associated with the element $a\in A$, and let  
    $$ M = \left( \begin{array}{ccc} 1 & 1 \cdots 1  \\ \vdots &\multirow{2}{*}M'\\ 
1 &  \\ 
 \end{array}  \right)$$
 the duality related to $\chi$. Since $\chi$ is a character we have $\sum_{b\in A} \chi_a(b) = 0.$
 Therefore,  $\sum_{i=1}^{r} M'_{i,j}=-1.$ Then, by Theorem~\ref{theorem:important} and applying Equation \ref{eq:EW-perp} with matrix 
   $$   \overline{M} = \left( \begin{array}{cc} 1 & r  \\ 1 & -1  \\ 
 \end{array}  \right),$$
  we obtain $HW_{C^M}(x_1,x_2) = \frac{1}{|C|} HW_C(x_1+rx_2,x_1-x_2).$
\end{proof}

\begin{proposition}\label{prop:Hamming2}
     Let $A=\{a_0,\dots,a_r\}$ be a group or a ring with $a_0=0$, and $\sim_H$ the equivalence relations given by the Hamming weight of the elements of $A$. Then
     $$
     HW_C(x_1,x_2)=W_C(x_1,x_2).
     $$
\end{proposition}
\begin{proof}
    Note that, given $\vc\in A^n$, we have that $N_2(\vc)=wt(\vc)$. Therefore, $N_1(c)=n-N_2(\vc)=n-wt(\vc)$. Hence, Equations \ref{eq:W_C} and \ref{eq:HW_C} coincide.
\end{proof}
We have that the weight enumerator of a code with respect to the equivalence relation given the Hamming weight is, in fact, the Hamming weight enumerator. Note that Equations \ref{eq:C*} and \ref{eq:HWC-perp} coincide.

In both examples, Examples \ref{ex:HammingF3} and Example \ref{ex:HammingZ4}, the duality $M$ is symmetric. However, propositions \ref{prop:Hamming1} and \ref{prop:Hamming2} are given for a general duality $M$, no necessarily symmetric. Next, we consider an example with a nonsymmetric duality.

 \begin{example} \label{firstexample} 
Let $A=\FF_4=\{0,1,\omega,\omega+1\}$, and consider the equivalence class that gives the Hamming weight; so that the alphabet has a partition into $A_1 = \{0\}$ and $A_2 = \{1,\omega,\omega^2\}.$ Consider the following duality on the finite field $\FF_4$:
$$ 
M= \left(\begin{array}{cccc}
 %M& 0 & 1 & \omega & 1+ \omega \\
  1 & 1 & 1 & 1 \\
  1 &-1 &1 &-1 \\
  1 &-1 &-1 &1 \\
  1 &1& -1& -1 \\
\end{array}\right),
$$
%\ \ \ \ \ \
% \begin{array}{c|rrrr}
% M^T & 0 & 1 & \omega & 1+ \omega \\
% \hline
% 0 & 1 & 1 & 1 & 1 \\
% 1 & 1 &-1 &-1 &1 \\
% \omega & 1 &1 &-1 &-1 \\
% 1+\omega & 1 &-1& 1& -1 \\
%\end{array} 
which is not symmetric.
Note that $\sum_{b \in A_2} M_{1,b}=\sum_{b \in A_2} M_{\omega,b} =\sum_{b \in A_2} M_{w^2,b} = -1$,
and $\sum_{b \in A_2} M_{0,b} =3$.
This gives that  $$ 
\overline{M}=\left( \begin{array}{cc} 1 & 3 \\ 1 &  -1 \end{array}  \right),$$ which gives the MacWilliams relation for the Hamming weight enumerator over $\FF_4$.
    
\end{example}

\subsection{Symmetric weight enumerators}\label{subsection:Symmetric}

We shall now show that the symmetric weight enumerator for additive codes over a group has MacWilliams relations.

\begin{definition}
Let $A=G$ be a finite abelian group. 
Define an equivalence relation on $G$ by $a \sim_S b $ if and only if $a= \pm b.$ Let $A_1, A_2,\dots,A_s$ be the equivalence classes of $\sim_S$. Let  $C$ be a code over $A$. Define the symmetric weight enumerator as 
\begin{equation} SW_C(x_1,x_2,\dots,x_s) = \sum_{\vc \in C} \prod_{i=1}^s x_i^{N_i(\vc)}. \end{equation}  
%$N_i(\vc) = | \{ j \ | \ c_j \in  A_i \}|.$ 
\end{definition}

\begin{example}
Let $A=\ZZ_4$. Then the equivalence classes of $A$ obtained from $\sim_S$ are $A_1=\{0\}$, $A_2=\{1,3\}$, $A_3=\{2\}$. Let $C=\{(0,0,0),(0,1,2),(0,2,0),(0,3,2),(2,0,0),(2,1,2),$ $(2,2,0),$ $(2,3,2)\}$.
Then,
$$
SW_C(x_1,x_2,x_3)=x_1^3+2x_1x_2x_3+2x_1^2x_2+2x_1x_3^3+x_2x_3^2.
$$
\end{example}

\begin{remark}
Note that if $A=\{a_0,\dots,a_r\}$ has characteristic $2$, then $A$ has $r+1$ classes, where $A_i=\{a_{i-1}\}$ for $i\in\{1,\dots,r+1\}$, and $SW_C(x_1,x_2,\dots,x_{r+1}) = cwe(x_1,x_2,\dots,x_{r+1})$.
\end{remark}

\begin{theorem}\label{theo:symm}
Let $M$ be a duality on the abelian group $G$.  Let $A_1,A_2, \dots, A_s$ be the equivalence classes related to $\sim_S$, where $a \sim_S b$ if and only if $a= \pm b.$  
Let  $\overline{M}_{A_i,A_j} = \sum_{b \in A_j}  M_{a,b}$,
then 
\begin{equation}
SW_{C^M} (x_1,x_2,\dots,x_{s})= \frac{1}{|C|}SW_C(\overline{M} \cdot (x_1,x_2,\dots,x_{s})).
\end{equation} 
\end{theorem} 
\begin{proof}

We shall use Theorem~\ref{theorem:important}.
We need to show that for all $i\in \{1,\dots,s\}$, we have that
$\sum_{b \in A_j} M_{a,b} = \sum_{b \in A_j} M_{a',b}$ if $a,a' \in A_i$. Note that $A_i=\{a,-a\}$, where either $a=-a$ or $a\not=-a$. If $A_i=\{a\}$, then it is clear. Assume now that $A_i=\{a,-a\}$ with $a\not=-a$. We only have to check that $\sum_{b \in A_j} M_{a,b}= \sum_{b \in A_j} M_{-a,b}$ for all $j\in\{1,\dots,s\}$.
If $A_j=\{b,-b\}$ then we have 
\begin{eqnarray*} \sum_{b \in A_j} M_{a,b} &=& M_{a,b} + M_{a,-b} \\  &=& \chi_a(b) + \chi_a(-b) = \chi_a(b) + (\chi_a(b))^{-1},
\end{eqnarray*} 
and   
\begin{eqnarray*}\sum_{b \in A_j} M_{-a,b} &=& M_{-a,b} + M_{-a,-b} \\ &=&  \chi_{-a}(b)  + \chi_{-a} (-b) \\ &=& (\chi_a(b))^{-1} + ((\chi_a(b))^{-1})^{-1} \\ &=& \chi_a(b) + (\chi_a(b))^{-1} .\end{eqnarray*} 
Finally, if $A_j=\{b\}$, $M_{a,b}=\chi_{a}(b)$ and $M_{-a,b}=(\chi_a(b))^{-1}=(\chi_a(-b))^{-1}=\chi_a(b)$. Then Theorem~\ref{theorem:important} gives the result.

\end{proof}

\begin{example}
For $\ZZ_4$, we have seen the MacWilliams relations when we consider the Hamming weight equivalence in Example \ref{ex:HammingZ4}. There is one other equivalence relation that gives MacWilliams relations.  That is, consider $\sim_S$ that gives the equivalence classes $\{0\}, \{2\}, \{1,3\}$, and
we get the following matrix for the MacWilliams relations
$$ \overline{M} = \left( \begin{array}{cccc} 1 & 2 & 1   \\ 1 & 0 & -1   \\ 1 & -2 & 1 \\
 \end{array}  \right).$$ 
\end{example}

What Theorem \ref{theo:symm} shows is that the symmetric weight enumerator has MacWilliams relations for any additive code with any duality.  Previously, it was only shown for codes over rings with the Euclidean inner-product, see \cite{Wood}.  

\subsection{$\lambda$ Weight Enumerator} \label   {subsection:lambda}

	Let $\lambda$ be a unit in the finite commutative  Frobenius ring $R$, with $\lambda^2=1.$  Partition the elements of $R$ by the equivalence relation
 $a \sim_\lambda b $ if and only if $a= \lambda b$ or $a=b$.  Note that if $\lambda=-1$, then both symmetric equivalences, $\sim_S$ and $\sim_\lambda$ coincide.

	\begin{definition}
Let $C$ be a code over a finite commutative Frobenius ring  $R$. Let $A_1,A_2,\dots,A_s$ be the equivalence classes under the equivalence relation $\sim_\lambda.$  The $\lambda$-weight enumerator is defined as 
\begin{equation}
\lambda W_C(x_1,\dots,x_s) = \sum_{\vc \in C} \prod_{i=0}^s x_i^{N_{i}(\vc)}.
\end{equation} 
%where $n_{i}(\vc) = | \{j \ | \ c_j \in A_i \} |. $
\end{definition}

	\begin{theorem} ({\bf MacWilliams Relations for $\lambda W$})
Let $M$ be the matrix that gives the MacWilliams relations for the complete weight enumerator over a chain ring $R$ as in Equation \ref{CCWE}.  
Define the matrix $\overline{M}$ by 
\begin{equation}
\overline{M}_{a,b} = \sum_{j \in A_b} M_{i,j} = \sum_{j \in A_b} \chi(ij) \end{equation} for some $i \in A_a.$  
Then
\begin{equation}
\lambda W_{C^\perp} (x_0,x_1,\dots,x_s) = \frac{1}{|C|} \lambda W_C(\overline{M} \cdot (x_0,x_1,\dots,x_s)).
\end{equation} 
\end{theorem} 
	\begin{proof} 
	
We need to show that this sum $\sum_{j \in A_b} \chi(ij)$ is the same for every $i \in A_a.$	
Let $a \in A_i$ and let $ A_j = \{ \alpha, \lambda \alpha \}$.  
Then $a A_j = \{ a\alpha, a \lambda \alpha\} $ and 
$a \lambda A_j = \{ a \lambda \alpha, a \lambda^2 \alpha \} = \{ a \lambda \alpha, a  \alpha \}.$  
Then $\sum_{j \in A_b} \chi(ij) = \sum_{j \in A_b} \chi(i'j)$ if $i,i' \in A_a.$ 
Then we have the result by Theorem~\ref{theorem:important}.  
\end{proof}

\begin{example}
 Consider the ring $\ZZ_8$, then there are $4$ units $\lambda$ with $\lambda^2=1,$ that is $1,3,5,$ and $7$.
 Then the equivalence classes are 
 $\{0 \}, \{1,3,5,7 \}, \{ 2,6 \}$ and
 $\{4 \}.$ In this case, the $\lambda$-weight enumerator coincides with the chain ideal weight enumerator.
For the ring $ \ZZ_{12}$, there are $4$ units $\lambda$ with $\lambda^2=1$,  that is $1,5,7,$ and $11$.
The equivalence classes are $\{0\},\{1,5,7,11\}, \{2,10\}, \{3,9\},\{4,8\},$ and $\{6\}$.  Note that in this case there is no chain ideal weight enumerator since the ring is not a chain ring.  
\end{example}

\subsection{Chain Ring Weight Enumerators} \label{subsection:CRW} 
Now we study two weight enumerators, $CRW$ and $\mathcal{OCRW}$ for codes over a chain ring $R$,  both of which are related to the order of elements. In the first case, $CRW$ is related to the order of each coordinate of the codewords in the code, while in the second case $\mathcal{CROW}$ is related to the order of the codewords in the code. The first one, helps us to study the Hamming weight of codes over a field $\FF_q$ obtained as the image under the Gray map of codes over a chain ring. The second one, gives the number of codewords of each order and it is closely related to the type. Both weight enumerators are also related to each other.

Let $R$ be a finite chain ring of index $e$, with maixmal ideal ${\mathfrak m} = \langle \gamma \rangle$. Recall that $\fa_i=\langle\gamma^{e-i}\rangle$ and $\fb_i = \fa_i \setminus \fa_{i-1}$ is the set of all elements in $R$ of order $\gamma^i$.  Also recall that $\fb_0,\fb_1,\fb_2,\dots,\fb_e$ form a partition of the ring $R$.  

% \begin{example} \label{primo} 
% Consider the chain ring $\ZZ_{16}$.   
% \end{example}

\begin{definition}
Let $A=R$ be a chain ring of index $e$ with maximal ideal ${\mathfrak m} = \langle \gamma \rangle $. For $a,a'\in R$, we define the equivalence relation $a\sim_o a'$ if $a$ and $a'$ have the same order in $R$. We have the equivalence classes $A_0,\dots,A_e$, where, for $i\in\{0,\dots,e\}$, $A_i=\fb_i$; that is all the elements of $A$ of order $\gamma^i$. Let $C$ be a code over the chain ring $R$. The chain ideal weight enumerator is defined as 
\begin{equation}
CRW_C(x_0,x_1,\dots,x_e) = \sum_{\vc \in C} \prod_{i=0}^e x_i^{N_{i}(\vc)},
\end{equation} 
where $N_{i}(\vc) = | \{j\in\{1,\dots,n\} \ | \ c_j \in \fb_i \} |. $
\end{definition}

\begin{example}\label{example:CRW}
Consider the chain ring $R=\ZZ_{16}$. Here, $\fb_0 = \{ 0 \}$, $\fb_1 = \{ 8 \}$, $\fb_2 = \{4,12 \}$, $\fb_3 = \{ 2,6,10,14 \}$, 
and $\fb_4 = \{1,3,5,7,9,11,13,15 \}$. The elements in $\fb_1$ have order $2^1=2$, the elements in $\fb_2$ have order $2^2=4$, the elements in $\fb_3$ have order $2^3=8$, and the elements in $\fb_4$ have order $2^4=16.$ 

Consider the code $C=\langle (1,8)\rangle$ over $R$. We have that 
\begin{align*}
    C=\{&(0,0),(1,8),(2,0),(3,8),(4,0),(5,8),(6,0),(7,8),\\
    &(8,0),(9,8),(10,0),(11,8),(12,0),(13,8),
(14,0),(15,8)\},
\end{align*}
and the chain ideal weight enumerator is
$$
CRW_C(x_0,x_1,x_2,x_3,x_4)=x_0^2+8x_1x_4+4x_0x_3+2x_0x_2+x_0x_1.
$$

\end{example}

Note that a field is a chain ring, namely the maximal ideal is $\{0 \}.$  Hence, for a field we have $\fb_0= \{ 0 \}$ and $\fb_1 = \FF \setminus \{ 0\}.$ Therefore, the chain ideal weight enumerator for finite fields is simply the Hamming weight enumerator.  We have the following statement.

\begin{proposition}
    Let $A=K$ be a field. Then, $e=1$ and 
    \begin{equation}
CRW_C(x_0,x_1) = W_C(x_0,x_1).
\end{equation} 
    
\end{proposition}

\begin{theorem} ({\bf MacWilliams Relations for $CRW$})
Let $M$ be the matrix that gives the MacWilliams relations for the complete weight enumerator over a chain ring $R$ as in Equation \ref{CCWE}.  
Define the $(e+1)\times (e+1)$ matrix  $\overline{M}$ with entries from $\CC$
where
\begin{equation}
\overline{M}_{a,b} = \sum_{j \in \fb_b} M_{i,j} = \sum_{j \in \fb_b} \chi(ij) \end{equation} for some $i \in \fb_a.$  
Then
\begin{equation}
CRW_{C^\perp} (x_0,x_1,\dots,x_e) = \frac{1}{|C|} CRW_C(\overline{M} \cdot (x_0,x_1,\dots,x_e)).
\end{equation} 
\end{theorem} 
\begin{proof}
We need to show that this sum $\sum_{j \in \fb_b} \chi(ij)$ is the same for every $i \in \fb_a.$

Consider the set $i \fb_b.$ By Lemma \ref{lemma:fafb}, $i=\gamma^{e-a} \eta$, for $\eta$ a unit in $R$. Also, from Lemma \ref{lemma:fafb}, all the elements in $\fb_b$ are of the form $\gamma^{e- b} \delta$ for $\delta$ a unit in $R$. Then, we have
\begin{eqnarray*} i \fb_b &=& \{ i \gamma^{e- b} \delta \ | \ \delta {\rm \ a \ unit \ } \} \\
 &=&  \{ \gamma^{e-a} \eta p^{e-b} \delta \ | \  \delta {\rm \ a \ unit \ } \}  \\
 &=&  \{ \gamma^{2e-a-b} \eta  \delta \ | \  \delta {\rm \ a \ unit \ } \}  \\
 &=&  \{ \gamma^{2e-a-b} \mu \ | \ \mu {\rm \ a \ unit \ } \} \\
 &=& \fb_{a+b -e}.
\end{eqnarray*} 
Therefore, $\sum_{j \in \fb_b} \chi(ij) = \sum_{j \in \fb_b} \chi(i'j)$ if $i,i' \in \fb_a.$  
\end{proof} 

Finally, for a code $C$ over a chain ring $R$, we relate the complete weight enumerator over a chain ring to the Hamming weight enumerator of its image under the Gray map given in Equation \ref{eq:GrayMapCarlet}.
\begin{theorem}\label{theo:PhiCRW}
 Let $C$ be a code over the chain ring $R$ of index $e$ and maximal ideal ${\mathfrak m}=\langle \gamma \rangle$ such that $R/{\mathfrak m}\cong \F_q$.  Then,
 $$
 W_{\Phi(C)}(x,y)=CRW_C(x^{q^{e-1}},y^{q^{e-1}},
 y^{(q-1)q^{e-2}}x^{q^{e-1}-(q-1)q^{e-2}},\dots,y^{(q-1)q^{e-2}}x^{q^{e-1}-(q-1)q^{e-2}})
 $$
\end{theorem}

\begin{proof}
Let $C$ be a code over the chain ring $R$, and consider the code $\Phi(C)$ over $\F_q^{q^{e-1}}$. Let $\vc=(c_1,\dots,c_n)\in C$, and $\Phi(\vc)=(\phi(c_1),\dots,\phi(c_n))$. By Equation \ref{eq:phi-isometry}, we have that $w(\Phi(\vc))= w_{Hom}(\vc)$; that is $\sum_{i=1}^n w(\phi(c_i))=\sum_{i=1}^n w_{Hom}(c_i)$. 

Let $i\in\{1,\dots,n\}$. By the definition of the homogeneus weight, we have that $w(\phi(c_i))=0$ if $c_i\in \fb_0$, $w(\phi(c_i))=q^{e-1}$ if $c_i\in \fb_1$
and $w(\phi(c_i))= (q-1)q^{e-2}$ otherwise. Since the length of $w(\phi(c_i))$ is $q^{e-1}$, the statement follows.
\end{proof}

\begin{example}
Consider the code $C$ given in Example \ref{example:CRW}. We have that, considering the matrix 
$$Y=\left(\begin{array}{cccccccc}
0&1&0&1&0&1&0&1\\
0&0&1&1&0&0&1&1\\
0&0&0&0&1&1&1&1
\end{array}
\right)$$
in the definition of the Gray map $\Phi:\ZZ_{16}\longrightarrow\ZZ_2^8$, we have that 
\begin{align*}
    \Phi(C)=\{&(0,0,0,0,0,0,0,0,0,0,0,0,0,0,0,0),
(0,1,0,1,0,1,0,1,1,1,1,1,1,1,1,1),\\
&(0,0,1,1,0,0,1,1,0,0,0,0,0,0,0,0),
(0,1,1,0,0,1,1,0,1,1,1,1,1,1,1,1),\\
&(0,0,0,0,1,1,1,1,0,0,0,0,0,0,0,0),
(0,1,0,1,1,0,1,0,1,1,1,1,1,1,1,1),\\
&(0,0,1,1,1,1,0,0,0,0,0,0,0,0,0,0),
(0,1,1,0,1,0,0,1,1,1,1,1,1,1,1,1),\\
&(1,1,1,1,1,1,1,1,0,0,0,0,0,0,0,0),
(1,0,1,0,1,0,1,0,1,1,1,1,1,1,1,1),\\
&(1,1,0,0,1,1,0,0,0,0,0,0,0,0,0,0),
(1,0,0,1,1,0,0,1,1,1,1,1,1,1,1,1),\\
&(1,1,1,1,0,0,0,0,0,0,0,0,0,0,0,0),
(1,0,1,0,0,1,0,1,1,1,1,1,1,1,1,1),\\
&(1,1,0,0,0,0,1,1,0,0,0,0,0,0,0,0),
(1,0,0,1,0,1,1,0,1,1,1,1,1,1,1,1)\},
\end{align*}

Applying Theorem \ref{theo:PhiCRW} to the polynomial $CRW_C(x_0,x_1,x_2,x_3,x_4)$ in Example \ref{example:CRW}, we obtain
\begin{align*}CRW_C(x^8,y^8,x^4y^4,x^4y^4,x^4y^4)&=x^{8\cdot2}+8y^8y^4x^4+4x^8y^2x^4+2x^8y^4x^4+y^8x^8\\
&=x^{16}+10y^{12}x^4+4x^{12}y^4+y^8x^8\\
&=W_{\Phi(C)}(x,y).
\end{align*}
\end{example}

%\subsection{Order Chain Rings Weight Enumerators}\label{subsection:OW}

Let $C$ be a linear code over $R$ of length $n$. Let $\fA_i$ and $\fB_i=\fA_i\setminus \fA_{i-1}$ be the subsets of $C$ defined in Section~\ref{section:chain} containing all the codewords of $C$ of order at most $\gamma^i$ and of order $\gamma^i$, respectively. Recall also that $\fB_0,\fB_1,\fB_2,\dots,\fB_e$ form a partition of the code $C$.

\begin{definition} Let $C$ be a code over the chain ring $R$ of index $e$ and maximal ideal ${\mathfrak m}=\langle \gamma \rangle$ such that $R/{\mathfrak m}\cong \F_q$, $q=p^m$, with $p$ prime.
    %Let $C$ be a code over the chain ring $R$ where $|R/\fm|$ has characteristic $p$ and $\fm=\langle \gamma \rangle$, where $\gamma$ has order $e$. 
    Define the order weight enumerator to be
    \begin{equation}
        {\mathcal {OCRW}}_C(z) = \sum_{i=0}^{e} A_z z_i
    \end{equation}
    where there are $A_i$ codewords of order $\gamma^i$ in the code $C$.
\end{definition}

\begin{theorem} \label{counter2} 
Let $C$ be a code over the chain ring $R$ of index $e$ and maximal ideal ${\mathfrak m}=\langle \gamma \rangle$ such that $R/{\mathfrak m}\cong \F_q$, $q=p^m$, with $p$ prime.
We have
    \begin{equation}
        {\mathcal {OCRW}}_C(z) = \sum_{i=0}^{e} |\fB_i| z_{i}.
    \end{equation}
   % where $i$ is the order of $\gamma^{e-i}.$ 
\end{theorem} 
\begin{proof}
    Follows from Theorem~\ref{autre}.   
\end{proof}

 Note that in this case we can not apply Theorem~\ref{theorem:important} to obtain the MacWilliam relations related to an equivalence class because the equivalence classes $\fB_0,\fB_1,\fB_2,\dots,\fB_e$ are not a partition of the alphabet.
 
\begin{example}
Consider the code $C=\langle (1,8)\rangle$ over $\Z_{16}$ given Example \ref{example:CRW}. Then 
\begin{align*}
\fB_0&=\{(0,0)\},    \\
\fB_1&=\{(8,0)\},\\
\fB_2&=\{(4,0),(12,0)\},\\
\fB_3&=\{(2,0),(6,0),(10,0),
(14,0)\}, \\
\fB_4&=\{(1,8),(3,8),(5,8),(7,8),(9,8),(11,8),(13,8),(15,8)\}.
\end{align*}

Then, we have that ${\mathcal {OCRW}}_C(z)=z_0+z_1+2z_2+4z_3+8z_{4}.$

\end{example}

Theorem~\ref{counter}  and Theorem~\ref{counter2} give that   ${\mathcal{OCRW}_C(z)}$ is completely determined by the type of the code.  Then 
Theorem~\ref{anyortho}  gives that type of $C^M$ is completely determined for any duality $M$.  Define the map $\vartheta$
 to be the following:
 \begin{equation} \label{sisi}
 \vartheta( {\mathcal{OCRW}}_C(z) ) = {\mathcal{OCRW}}_{C^M}(z).  
 \end{equation}

\begin{definition} \label{psis} 
    Define \begin{equation} \label{littlepsi} 
    \psi\left(\prod_{i=0}^e x_i^{N_i(\vc)}\right) = z_{\max \{ i \ | \ N_i(c) \neq 0 \}} . \end{equation} 
    Then define 
    \begin{equation}\label{bigpsi}  \Psi\left(\sum A_{\gamma_0,\gamma_1,\dots, \gamma_e} \prod_{i=0}^e x_i^{\gamma_i} \right)= \sum A_{\gamma_0,\gamma_1,\dots, \gamma_e} \psi\left(\prod_{i=0}^e x_i^{\gamma_i}\right).  \end{equation} 
\end{definition}

\begin{theorem} Let $C$ be a linear code over the chain ring $R$. Then 
\begin{equation} \Psi(CRW_C(x_0,x_1,\dots,x_e)) = {\mathcal{OCRW}}_C(z). \end{equation} 
\end{theorem} 
\begin{proof}
    Follows from Lemma~\ref{orders} and  Definition~\ref{psis}. 
\end{proof}

Let $\Omega$ be the map defined by 
\begin{equation} \Omega_M(CRW_C(x_0,x_1,\dots,x_e)) =  \frac{1}{|C|} CRW_C(\overline{M} \cdot (x_0,x_1,\dots,x_e))\end{equation}

Given the results in this section we have that the following diagram commutes.

\[ \begin{tikzcd}
CRW_C (z)\arrow{r}{\Omega_M} \arrow[swap]{d}{\Psi} & CRW_{C^M}(z)  \arrow{d}{\Psi} \\%
\mathcal{OCRW}_{C}(z) \arrow{r}{\vartheta}& \mathcal{OCRW}_{C^M}(z)
\end{tikzcd}
\]

We can also use the results to get a result about the ideals in a chain ring.

\section{Conclusions}

Codes can be defined over any alphabet $A$.  If we want the full force of the MacWilliams relations which determine the weight enumerator of the dual code from the original code, we restrict ourselves to additive codes over a finite abelian group or linear codes over a Frobenius ring. For codes over the finite field $\FF_q$, we can consider both additives codes or linear codes over the underlying group or ring of $\FF_q$. In the case of additive codes there are numerous different dualities that can be used to define an orthogonal and for linear codes we can use the standard Euclidean inner-product.  
In all of these cases, various weight enumerators can be defined and MacWilliams relations can be given for these weight enumerators that can be deduced from the general MacWilliams relations for the complete weight enumerator.  

The given results on the MacWilliams relations are foundational results for algebraic coding theory as there are a wide variety of applications within the theory that are dependent upon these results.  In this paper, we have added to this literature giving interesting new weight enumerators and describing how the MacWilliams relations can be determined for them. For example, if we have an equivalence relation defined on the alphabet $A$, we can then define a new weight enumerator intimately related to this relation.
We show that if it satisfies certain algebraic conditions then MacWilliams relations can be produced for them.  In particular, we have defined a new weight enumerator on codes over a finite chain ring defined by an algebraic equivalence relation that gets to the very core of understanding the structure of codes.


\begin{thebibliography}{99}

\bibitem{z2z4}
J. Borges,
C. Fern\'andez-C\'ordoba, J. Pujol, J. Rif\`a, M. Villanueva, 
{$\ZZ_2\ZZ_4$-linear codes}, 
Cham: Springer (ISBN 978-3-031-05440-2/hbk; 978-3-031-05441-9/ebook), 2022.


\bibitem {Delsarte} 
P. Delsarte, 
{ An algebraic approach to the association schemes of coding theory}, 
{\em Philips Research Reports}. Supplements 10. Ann Arbor, MI: Historical Jrl. vi, {\bf 97},   1973.


\bibitem{mybook} S.T. Dougherty, {\em Algebraic Coding Theory over Finite Commutative Rings},  Springer Briefs in Mathematics. Springer, 2017.

\bibitem{bigduality} S.T. Dougherty, {  Dualities for Codes over Finite Abelian Groups}, {\em Advance in Mathematics of Communication}, {\bf 18}, no. 6, pp. 1827-1841, 2024.


\bibitem{open} S. T. Dougherty, J.-L. Kim, P. Sol\'e,  Open problems in coding theory, {\em Contemp. Math.}, {\bf 634}, pp. 79 - 99, 2015.
\bibitem{Involve}  S. T. Dougherty, S. Myers,  Orthogonality from Group Characters,  {\em Involve}, {\bf 14}, no. 14,  2021. 

\bibitem{doughertyeven} S.T. Dougherty, S. Sahinkaya, Dualities over the cross product of the cyclic groups of order $2$, {\em Advance in Math. of Communication},  {\bf 18}, no. 5, pp. 1531-1546, 2024.


\bibitem{NariLee} S.T. Dougherty, J. -L. Kim, N. Lee, Additive self-dual codes over finite fields of even order, {\em Bull. Korean Math. Soc}, {\bf 55}, no 2, pp. 341- 357, 2018.  

\bibitem{non-comm} S. T. Dougherty, A. Leroy, Self-dual codes over non-commutative Frobenius rings,   {\em Applicable Algebra in Engineering, Communication and Computing}, {\bf 27}, no. 3, pp. 185-203, 2016,  DOI 10.1007/s00200-015-0277-0. 


\bibitem{Heide} H. Gluesing-Luerssen, 
Fourier-reflexive partitions and MacWilliams identities for additive codes, {\em 
Des. Codes Cryptog.}, {\bf 75}, no. 3, 543-563, 2015.


%\bibitem{Rings4} B. Fine,  Classification of Finite Rings of Order $p^2$,
%Classification of finite rings of order $p^2$, Math. Mag. {\bf 66},   no. 4, 248-252, 1993. 

\bibitem{GrayIsometry} M. Greferath and S. E. Schmidt, Gray isometries for finite  chain rings and a nonlinear ternary {$(36,312,15)$} code, {\em IEEE Transactions on Information Theory}, vol. 45, no. 7, pp. 2522-2524, 1999.

\bibitem{Huffman} W.C. Huffman, V. Pless, 
{\em Fundamentals of error-correcting codes}, 
Cambridge: Cambridge University Press (ISBN 0-521-78280-5/hbk). xvii, 646 p. (2003).

\bibitem{Jitman} J. Jitman and P. Udomkavanich, The Gray image of codes over finite chain rings, {\em Int J Contemp Math Sci}, {\bf 5}, pp. 449-458, 2010.

\bibitem{NOMAC} M. R. Julian, No MacWilliams duality for codes over nonabelian groups, {\em JACODESMATH}, {\bf 5}, Issue 1, 2018.

\bibitem{MacD} B. R. McDonald, {\em Finite Rings with Identity}, Dekker, New York (1974). 

\bibitem{thesis} F.J. MacWilliams,  {\em 
Combinatorical Problems of Elementry Abelian groups}, 
Thesis (Ph.D.)–Radcliffe College,
ProQuest LLC, Ann Arbor, MI, 1962.



\bibitem{Mac2}
F.J.  MacWilliams,  A theorem on the distribution of weights in a systematic code, {\em Bell System Tech. J.},  {\bf 42}, pp. 79 - 94, 1963.

\bibitem{SalageanChain} G. Norton and A. Sălăgean, On the Structure of Linear and Cyclic Codes over a Finite Chain Ring, {\em Applicable Algebra in Engineering, Communication and Computing}, {\bf 10}, pp. 489-506, 2000.
 
%\bibitem{OEIS}  N. J. A. Sloane, The On-Line Encyclopedia of Integer Sequences, http://oeis.org.
 
\bibitem{Wood}  J. Wood,
{Duality for modules over finite rings and applications to coding
theory}, {\em Amer. J. Math.},  {\bf  121}, no. 3,  pp. 555 - 575, 1999.

\bibitem{Wood2} J. Wood, 
Foundations of linear codes defined over finite modules: the extension theorem and the MacWilliams identities, Codes over rings, 124-190,{\em Ser. Coding Theory Cryptol.}, {\bf 6}, 
World Scientific Publishing Co. Pte. Ltd., Hackensack, NJ, 2009.
\end{thebibliography}
\end{document}